# Crafting Tomorrow's Evaluations: Assessment Design Strategies in the Era of Generative AI

Rajan Kadel
Melbourne Institute of Technology
rkadel@mit.edu.au

Bhupesh Kumar Mishra
University of Hull, UK
bhupesh.mishra@hull.ac.uk

Samar Shailendra
Melbourne Institute of Technology
sshailendra@mit.edu.au

Samia Abid
University of Hull, UK
samia.abid-2022@hull.ac.uk

Maneeha Rani
University of Hull, UK
m.rani3-2022@hull.ac.uk

Shiva Prasad Mahato
Khwopa Engineering College, Nepal
sp.mahato@khec.edu.np

*Abstract*— In recent years, no other technology has revolutionised our life as Generative Artificial Intelligence (GenAI). GenAI has gained the attention of a myriad of users in almost every profession. Its advancement has had an intense impact on education, significantly disrupting the assessment design and evaluation methodologies. Despite the potential benefits and possibilities of GenAI in the education sector, there are several concerns primarily centred around academic integrity, authenticity, equity of access, assessment evaluation methodology, and feedback. Consequently, academia is encountering challenges in assessment design that are essential to retaining academic integrity in the age of GenAI. In this article, we discuss the challenges, and opportunities that need to be addressed for the assessment design and evaluation. The article also highlights the importance of clear policy about the usage of GenAI in completing assessment tasks, and also in design approaches to ensure academic integrity and subject learning. Additionally, this article also provides assessment categorisation based on the use of GenAI to cultivate students' and academic professionals' knowledge. It also provides information on the skills necessary to formulate and articulate problems and evaluate the task, enabling students and academics to effectively utilise GenAI tools.

## I. INTRODUCTION

The key challenges in the education ecosystem by the intro- duction of Generative Artificial Intelligence (GenAI) assessment design are threefold: Ensure learning; maintain academic integrity; and gain knowledge and skills about GenAI tools and technologies. Therefore, the design of assessments in the era of GenAI should ensure a balance among all three outcomes. Many traditional assessments can be completed by GenAI applications with limited effort and knowledge. Thus, there is a need for proper assessment design to ensure academic integrity and student learning in the age of GenAI [1]. Moreover, the integration of GenAI in the assessments is essential for preparing future graduates to:

- Understand GenAI technologies, associated opportunities, and challenges,
- Apply GenAI tools responsibly and efficiently which is crucial to enhancing both learning & productivity, and
- Equip students with skills and knowledge to use emerging GenAI applications.

According to the report "Assessment reform for the age of artificial intelligence" by Tertiary Education Quality and Standards Agency (TEQSA) [2], the two principles to be ensured during assessment design in the era of GenAI are:

- Students should be able to use GenAI tools ethically and actively by recognising the ethics, limitations, biases, and implications of AI. The assessments need to consider both positive and negative risks posed by GenAI.

- Ensure student learning by using a variety of inclusive and contextualised approaches on assessments.

To achieve the aforementioned principles, it is responsibility of academics to ensure that assessments are designed to provide opportunities to students for learning on the subject matter and GenAI tools by using proper assessment design strategies.

The primary contributions of this paper include assessment classification, assessment design process and evaluation strategies ensuring that the students' learning outcomes are preserved in the era of GenAI.

The outline of the paper is as follows: Section II provides a brief review of assessment design and evaluation in the era of GenAI. The proposed assessment categorisation and how to ensure learning is described in Section III. Authentic assessment design and evaluation strategies for the proposed assessments are discussed in Section IV. Finally, Section V presents the concluding remarks and the future directions.

## II. RELATED WORKS

The assessment design process has been an area of de- bate since the inception of GenAI [3]. The researchers have contributed to the assessment design process by emphasising the ethical considerations and challenges of using ChatGPT-3 in academic environment [4]. The authors have emphasised ethical challenges associated with integrating ChatGPT-3 in education, particularly concerning plagiarism, academic integrity, and the potential for misuse. This research has argued that ethical leadership and character development are required to support the responsible use of AI technologies, highlighting the importance of fostering a supportive learning environment. These insights provide valuable guidance for designing and implementing ChatGPT-3 in educational settings. A frame- work has been developed for re-designing writing assignment assessments in the era of LLMs like ChatGPT and GPT-4, Bard, LLaMA, and BLOOM [5]. The proposed framework includes six dimensions: purpose, function and focus, grading criteria, modes, authenticity, and administration, all aimed at addressing the challenges posed by GenAI in the assessment design of writing assignments. The practical solutions provided for educators include implications for crafting course learning goals with GenAI, with criteria of accuracy, precision, relevance, depth, balance, and logic. These solutions serve as a stimulus for the dimensions of the framework and were used in workshops with educators, helping them to redesign their writing assignments [6].

A fine-tuning assessment strategy in engineering education to combat the risks of ChatGPT, particularly in online quizzes, and recommends terminating high-risk formats [7]. This work has provided practical integration suggestions including the use of ChatGPT for concept confirmation in non- assessed quizzes, embracing flipped assessments based on critical thinking, and vigilantly introducing the tool in areas where academic integrity risks are minimal. An investigation critically highlights the limitations of traditional assessment methods, including their onerous nature, lack of adaptability, and inauthenticity. The authors suggest that integrating AI into assessment practices can address these issues. It recognises the constraints of both traditional and GenAI approaches and urges to improve the ongoing assessment practices [8].

The authors in [9] have discussed innovative assessment strategies for addressing GenAI challenges in education, such as prioritising process over product, using oral assessments, employing AI detection tools, and promoting critical thinking. These strategies can be used to enhance both traditional and online evaluations. An investigation has been conducted employing a survey-based method to evaluate the impact of GenAI tools on assessment practices in higher education setting. The study utilised the survey-based approach to gather and analyse data, contributing to the ongoing discourse on assessment reform and the integration of technology in applied computing environment. According to [10], a valuable exploration of AI chatbots has used to improve learning outcomes in a classroom. Teachers utilise AI chatbots for the development of assessments and evaluation of their outcomes to critically analyse the understanding of complex topics by the students. This method provides visibility into what students understand and which areas need attention. 'Smart Grading' is a software that utilises GenAI to score the text-based answers in educational evaluation and behavioural studies [11]. This novel method offers a valuable tool for educators and researchers by integrating large language models and customised options. The software extends a flexible and effective solution for evaluating extensive amounts of text-based answers.

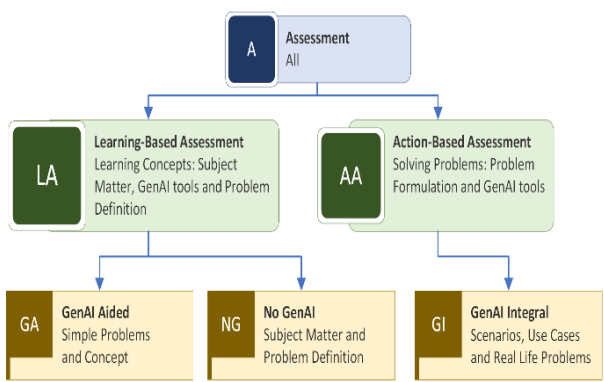

Fig. 1: Assessment classification in the era of GenAI.

## III. LEARNING ASSESSMENT CLASSIFICATION

In the era of GenAI, the assessments need to ensure the proper use and understanding of GenAI tools, the learning on the subject matter, and academic integrity. Figure 1 illustrates assessment categories in the era of GenAI where the assessments are divided in two types: **Learning-based Assessments** and **Action-based Assessments**. Learning-based Assessments are designed for learning subject matter and problem articulation. On the other hand, Action-based Assessments are designed to check the application of knowledge and skills on subject matter and problem articulation using GenAI tools.

### A. Learning-based Assessments

These assessments emphasise measuring and recalling factual concepts particularly inclined towards the instructor's choices. These assessments are mainly offered during the early stages of the program/subject while students are learning the fundamentals, problem formulation and skills on the use of GenAI tools. These assessments are further divided into two types: Assessments with no GenAI, and Assessments with GenAI aided mode.

*Assessments with No GenAI:* These assessments are designed to ensure learning the subject matter concept, and problem formulation. Some examples of such assessments are: In-person unseen examinations, class tests, online tests, vivas, laboratories and practical tests and discussion-based assessments. These assessments are carried out in both supervised and semi-supervised modes. In supervised or proctored assessments, examiners can observe students as they independently perform tasks, without the use of GenAI tools or any other assistance. However, students can use any tools including GenAI tools for learning before the assessment. Some specific tasks such as practical examination, oral presentation, and simulation-based tasks can be performed in semi-supervised settings. These assessment methods are contemporary methods of assessment to ensure students' learning.

*Assessments with GenAI Aided Mode:* These assessments are structured to help students grasp the fundamental use of GenAI tools. Examples may include, structuring and drafting content; supporting as a tutor; testing code; providing feedback on content, etc. In this type of assessment, academics may also introduce discipline-based GenAI tools. These types of assessments highlight the importance of students being aware of both the capabilities and limitations associated with GenAI.

### B. Action-based Assessments

These assessments are mainly offered during the advanced stages of the program/subject while students are applying their skills and knowledge to solve the given real-world problems or tasks. These assessments focus on the application of knowledge using any tools including GenAI where students demonstrate the application of knowledge. In these assessments, students can demonstrate their knowledge and skills in a useful way by solving real-life complex applications. At the same time, it also offers direct evidence and observability for the evaluators. The Action-based Assessments are adaptive and student-focused. They are designed in such a way that students need to demonstrate knowledge and its application by solving comprehensive complex problems. Therefore, structure, guidance, and rubric are the key elements for their design. To meet the objectives of Action-based Assessments, the key steps for design are:

*Identifying competencies:* In this step, competencies to be assessed are determined by the academics. The competencies to be assessed depend on students' knowledge, skills, learning requirements, and qualification level. Assessments should utilise action verbs such as design, analyse, present, or solve, etc to frame the expected actions.

*Choosing suitable tasks:* The tasks should mirror real-world situations where learners will apply their skills. The complexity level of the tasks should consider learner experience and qualification level. Additionally, the tasks should offer flexibility on different learning styles to accommodate diverse approaches or solutions. The competencies mentioned in the previous step should be clearly identified by academics. The assessment guidelines should indicate how students meet these competencies.

*Setting up criteria:* In this step, academics need to set up key criteria to measure the performance of tasks(s) in the previous step. The key in this step is to break down the task into actionable steps and focus on observable behaviours. Then, the criteria need to indicate what excellent, good, or bad performance on these actionable steps looks like. It is helpful and more productive if the competencies identification, task(s) selection, and measurement criteria mapping out are performed during assessment creation.

*Creating the rubric:* In this step, academics need to clearly describe instructions, resources, and evaluation rules for the performance of the task(s).

*Refine Assessment:* In this step, the assessment designer collects feedback from learners and educators on the assessments. Subsequently, the assessment is revised considering all design aspects (competency identification, task selection, measurement criteria and rubric formation) using the feed- back received.

*GenAI Integral Mode:* In this mode, students apply their skills and knowledge in the discipline areas, problem formulation skills, and GenAI tools (take advantage of) to solve complex real-life problems either in a group or individually. These tasks provide opportunities for students to work appropriately with each other and GenAI tools. Some examples of these types of tasks using GenAI in an integral model include drafting and structuring content; generating ideas; comparing content (GenAI generated and human-generated); creating content in particular styles; producing summaries; analysing content; re-framing content; researching and seeking answers; creating artwork (images, audio and videos); playing a Socrative role and engaging in a conversational discussion; developing code; translating content; generating initial content to be critiqued by students. The use of GenAI in assessments introduces several challenges that may require restructuring assessment design and evaluation strategies.

## IV. ASSESSMENT DESIGN AND EVALUATION STRATEGIES

This section presents strategies for assessment design and evaluation for the assessment types introduced in Section III.

### A. Assessment Design Strategies

As GenAI is revolutionising the whole education industry, there is a need for an assessment plan for a whole program before applying this strategy to ensure students learning on the theoretical concepts and GenAI tools. Figure 2 illustrates the proposed process of assessment design that involves the following steps and strategies:

*Step 1- Determine the type of assessment:* In this step, the types of assessments used in the program and subject level are planned in such a way that assessments are aligned with course and learning outcomes i.e. the assessment completion ensures that the learner has met the learning outcomes and grasped the essential skills and knowledge. This requires that there is a mix of Learning-based and Action-based Assessments in course curriculum. The key considerations while designing these assessment types are:

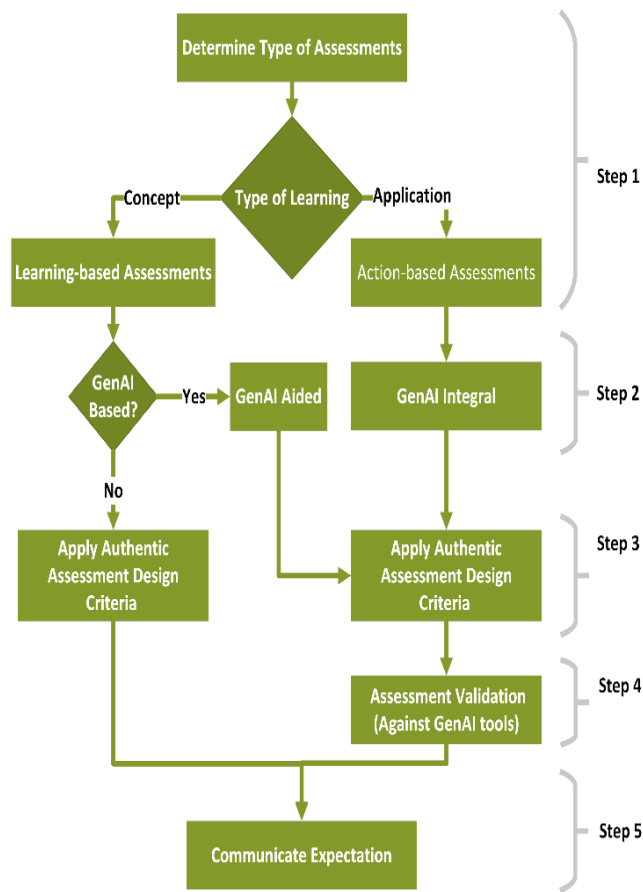

Fig. 2: Proposed assessment design process.

- There should be enough Learning-based Assessments to ensure that students are learning the concepts, GenAI tools, and problem-definition skills. This type of assessment needs to be selected in the early stage of the program or the unit so that students have learned the concepts before using the GenAI tools. Problem articulation is one of the key skills students should learn before being exposed to GenAI tools [12]. Therefore, Learning-based Assessments should concentrate on acquiring skills in formulating problems. Additionally, there should be enough student exposure to GenAI tools either inside the unit or entirely in a new unit to ensure that students understand their usage, benefits, and limitations.

- The Action-based Assessments always follow Learning- based Assessments. The Action-based Assessments are de- signed to check the student's ability to apply knowledge and skills using GenAI tools.

- The tasks used in both of these assessments need to apply authentic design techniques to ensure that

students apply critical thinking, creativity, and problem-solving skills to real-world scenarios.

***Step 2- Determine the use of GenAI:*** The clarity on the use of GenAI in both types of assessment is critical for students and academics. In the assessment plan for the course, there should be a clear instruction on the use of GenAI (on what level: No GenAI, GenAI Aided, or GenAI Integral) in each assessment. This is crucial for academics during assessment task selection, incorporating GenAI tools in assessment design and planning how to ensure the academic integrity of the task. Similarly, this is crucial that students are provided with clear guidelines about the usage of GenAI for completing the assessment task.

***Step 3- Apply Authentic Assessment Design:*** This step applies to all assessments to enhance learning outcomes and maintain academic integrity in the age of GenAI. The assessment design method needs to take a comprehensive approach valuing all key graduate attributes such as critical thinking, life-long learning, problem-solving skills, collaboration, and communication, in conjunction with their disciplinary knowledge, skills, and practical application. The key strategies that can be applied during authentic assessment design are as follows:

- Design assessment tasks prompt students to utilise critical thinking, creativity, and problem-solving abilities using practical and real-world scenarios.

- Include interactive assessment tasks that enable students to apply their knowledge into action and demonstrate their competency concretely through practical experiences.

- Create group assessment tasks where the academics can observe and assess the process of collaboration, communication, and interpersonal skills. Follow ***Process over Product (PoP)*** strategy i.e., the emphasis should be more on the process of creating the Artefact than on the final Artefact itself.

- Include assessment tasks where students demonstrate their ability to exercise evaluative judgement of the peer's or published works.

- Offer diverse forms of assessments or multi-modal assessments to ensure opportunities for students to demonstrate their learning and ensure academic integrity.

- Include context-specific/personalised design problems including problem-based or project-based tasks in the assessments where students can apply their learning and skill.

- Incorporate individual or team-based in-class assessments including quizzes, live polls, tests, concept maps, short written tasks, or oral presentations to observe students learning and participation.

***Step 4- Validate Assessments Against GenAI tools:*** This step is critical to ensure that students cannot complete the assessment tasks by only using GenAI tools, without proper learning or application of knowledge and skill. As some Learning-based Assessments and all Action-based Assessments are open to GenAI tools. Therefore, the compliance of learning and application of knowledge and skills is crucial to maintaining academic integrity in these types of assessments. The key strategies that can be applied during validation of assessments against GenAI tools are:

- Assessment tasks need to focus on the process or continuous assessment rather than the final Artefact in both individual and group assessments. Academics need to validate the assessment rubric before release to ensure that the process can be measured. The assessment can be designed in staged assessments covering either one semester or multiple semesters. More weighting should be given to the process involved than the actual final product. This strategy in assessment design ensures that academics have better insight into students learning and also fosters students' learning opportunities.

- Design and include specific assessment tasks incorporating GenAI tools to understand the capability and shortcomings associated with the tools. These tasks can be in various formats. Here is an example.

    ∗ Student are asked to utilise GenAI tools to obtain the solutions for the provided problems or tasks.

    ∗ Check the factual information on the output against various types of available sources.

    ∗ Critique the output from the GenAI tools with peer-reviewed papers, books, and course content or the given rubric in the assessment.

    ∗ Students are asked to check the quality of the output and rank them according to the outcomes of two previous steps.

    ∗ Write a personal reflection on the process.

- Broaden the array of assessments for different evaluation formats (video-based, portfolio, oral presentation, self and peer reflection, oral interview, peer-interview or role play, etc.,) to ensure student learning and academic integrity.

- Incorporate authentic, context-specific, or in-class assignments as discussed in step 3. Moreover, the assessments should be more focused on continuous learning and performance improvement rather than just providing a discrete snapshot of the student's performance.

Overall, check all assessment tasks against various GenAI tools and make sure that the tasks are designed such that their completion using GenAI tools will not compromise the requisite learning for the students. If it can be completed by GenAI tools without any effort by students, make necessary changes and iterate the process until satisfied.

**Step 5- Communicate Expectation:** This step is critical to bring clarity and setting the right expectations for all stakeholders. The institutions need to clearly communicate their willingness to adopt the GenAI in their strategy documents. Accordingly, the academic staff and the students should have a clear rationale behind overall assessments at the course level as well as the subject level. In each assessment task, the academic staff should clearly articulate the type and objective of the assessment, and specify the expectation in terms of using the GenAI in the assessment. The clear communication will ensure academic integrity and help to achieve the learning objectives.

## A. Assessment Evaluation Strategies

This is important to identify assessment evaluation strategies in the age of GenAI as the assessments are changing. Based upon the types of assessment as discussed in Section III, the key assessment evaluation strategies comprise a) Traditional approach for Learning-based Assessments; and b) Evaluation automation using GenAI for Action-based Assessments.

The traditional evaluation approach uses methods such as proctored assessment, evaluation by the educator, and peer review-based strategies for the evaluation and measurement of the learning item. On the other hand, it is important to understand and design effective evaluation strategies for Action-based Assessments. One of the important criteria is that these assessments should be designed such that they can be evaluated using the GenAI-based tools. This also implies that either such GenAI-based tools should exist or need to be developed to carry out the assessment evaluation.

Action-based Assessment methods (AbAM) [13] is an innovative approach that also allows evaluation for a virtual learning environment where either the experts are too costly to hire to provide feedback to learners or risky in particularly critical situations. AbAM uses a formative assessment method, which focuses on what actions the learner performs and how. This assessment approach allows learners to perform complete operations and observe consequences within system limitations instead of just predefined action choices. Learner actions are recorded, processed, and compared to experts' actions for generating formative feedback reports while interacting with virtual training environments. This method enables learners to learn from mistakes and repeat assessments until success.

## V. CONCLUSION AND FUTURE DIRECTIONS

In conclusion, this paper has explored the profound impact of GenAI on assessment design and evaluation methodologies. We have discussed the challenges and opportunities presented by GenAI, highlighting the need for innovative strategies to ensure academic integrity, and authenticity in assessment design and evaluation. The paper proposes assessment categorisation to ensure subject matter learning as well as enable the students to apply this knowledge and skills in the era of GenAI. This approach ensures that GenAI tools are used effectively and ethically in the educational context. Furthermore, our discussion on assessment design and evaluation strategies provides valuable insights for adoption of GenAI.

Looking ahead, we believe that the integration of GenAI in education will continue to evolve. While this paper offers broader guidelines and methodology, it's crucial to validate the approach through implementation and integrate the insights to refine the process. We hope that our work serves as a stepping stone for future assessment design in this exciting and rapidly evolving field of GenAI.